\documentclass[a4paper,11pt]{article}
\usepackage{jinstpub} 
\usepackage{graphicx}
\usepackage{subcaption}
\usepackage{tikz}
\usepackage[percent]{overpic}
\usepackage[font=small,labelfont=bf]{caption}

\usepackage{amssymb}
\proceeding{26$^{\text{th}}$ International Workshop on Radiation Imaging Detectors\\
   6--10 July 2025\\
  Bratislava, Slovakia}

\title{\boldmath Design and Evaluation of CZT-based Micro-activity Dose Calibrator for TAT Application using Monte Carlo Simulation}

\author[a]{Seoyun Jang,}
\author[b]{Robin Peter,}
\author[b]{Biswajit Das,}
\author[b]{Youngho Seo,}
\author[a,1]{Gyuseong Cho\note{Corresponding author.}}

\affiliation[a]{Department of Nuclear \& Quantum Engineering, Korea Advanced Institute of Science and Technology,\\
291 Daehak-ro, Daejeon, Republic of Korea}
\affiliation[b]{Department of Radiology and Biomedical Imaging, University of California, San Francisco,\\
185 Berry Street, San Francisco, USA}
\emailAdd{gscho1@kaist.ac.kr}

\abstract{
A novel CZT-based micro-activity dose calibrator has been designed via Monte Carlo simulation (GATE) to accurately measure low-level activity $^{225}$Ac for targeted alpha therapy (TAT) application.
Because even small overdoses in TAT can induce severe local toxicity, activities in the microcurie down to nanocurie regime are often required, and accurate activity measurement by dose calibrators is a priori to safe and effective treatment.
Standard dose calibrators, or high-pressurized gas-filled ionization chambers, are not suitable in this range due to limited sensitivity and lack of energy discrimination.
To address this, we designed a CZT-based micro-activity calibrator that (i) adopts a box-shaped well geometry to obtain higher solid-angle coverage and (ii) applies time-coincident pixel-level clustering to recover full-energy-peak net counts otherwise lost to multi-site Compton scattering.
Using GATE, serial dilutions from $1.0~\mu$Ci down to $10^{-6}~\mu$Ci were simulated, activities were reconstructed from the 218 and 440~keV gamma peaks and performance was compared against a NaI(Tl) well counter which serves as an alternative to standard dose calibrator.
Across six orders of magnitude, the CZT-based micro-activity dose calibrator exhibited near-unity linearity (slope $m=0.9934$) with percent-level bias, whereas the NaI(Tl) counter showed systematic under-response under identical conditions.
These results indicate that a CZT-based approach can provide accurate low-activity quantification for TAT, motivating forthcoming hardware validation.}


\begin{document}
\maketitle
\flushbottom

\section{Introduction}
\label{sec:intro}

Targeted alpha therapy (TAT) is a promising cancer treatment with active clinical and preclinical research ongoing. TAT provides highly specific tumor cell killing while minimizing damage to surrounding normal tissues~\cite{Intro1,Intro2}. 
This is possible by the physical properties of $\alpha$-particles, which exhibit high linear energy transfer (LET; $\sim$100~keV/$\mu$m) and short ranges (50--100~$\mu$m in tissue)~\cite{Intro3}. 
These properties mean that only a few $\alpha$-particle hits are often sufficient to kill a cell, since $\alpha$-induced DNA double-strand breaks are hard to repair and cytotoxicity shows minimal oxygen dependence~\cite{Intro4,Intro5}. 
Consequently, the cytotoxicity of $\alpha$-particles is considered extremely potent, with cell death achievable from only a single or a few $\alpha$-particle traversals of the cell nucleus~\cite{Intro6}.

Because even small overdoses in TAT can induce severe local toxicity in nearby healthy tissues, $\alpha$-emitting radionuclides are typically administered in small, fractionated doses~\cite{Intro7,Intro8,Intro9}. 
Among $\alpha$-emitters investigated for TAT, $^{225}$Ac is considered promising and has been most widely studied due to its physical properties of high delivery dose (four $\alpha$-particles emitted in its decay chain) and its half-life of 9.9 days~\cite{Intro2}.
For metastatic prostate cancer using PSMA-targeted agents (e.g., [\textsuperscript{225}Ac]Ac-PSMA-617 or -PSMA-I\&T), typical clinical activities are 50--100~kBq/kg per cycle, while preclinical mouse studies often use $\sim$11.1~kBq/mouse~\cite{Intro10,Intro11,Intro12,Intro13}. Beyond these ranges, preclinical work has also explored $^{225}$Ac at microcurie and even nanocurie levels, underscoring the need for accurate quantification in ultra-low-activity settings~\cite{Intro21}.

However, standard dose calibrators, or high-pressurized argon-filled ionization chambers, are not well suited to such low-activity measurements required for TAT.
Dose calibrators are devices used to quantify the activity of radiopharmaceuticals before they are administered to patients in nuclear medicine.
Standard dose calibrators provide accurate, linear measurements within $\pm$5--10\% for most radionuclides for relatively high activities, ranging from $\sim$37~kBq (1~$\mu$Ci) to 3.7~GBq (100~mCi)~\cite{Intro19}. 
In contrast, preclinical studies often demand $\sim$1\% measurement accuracy in the sub-microcurie range, which standard ionization chambers do not reliably achieve. 
Below 37~kBq (1~$\mu$Ci), the signal-to-noise ratio decreases substantially, making measurements unreliable and often necessitating dilution procedures, which may introduce additional human error ~\cite{Intro20}.

Accordingly, for low-activity measurements, NaI(Tl) well counters are widely adopted in both clinical and preclinical settings. 
A NaI(Tl) well counter employs a thallium-doped sodium iodide scintillation crystal coupled to a photomultiplier tube (PMT). 
It offers high detection efficiency due to the material properties of NaI ($\rho=3.67~\mathrm{g/cm^3}$, $Z_{\mathrm{eff}}\approx50$) and well-type geometry. 
Nevertheless, their limitations are significant: NaI(Tl) exhibits poor energy resolution ($\approx$6--9\% at 662~keV), and at higher $\gamma$ energies ($>250$~keV) Compton scattering dominates over photoelectric absorption, degrading spectral quality and reducing peak-to-total ratios. 
These drawbacks constrain quantitative accuracy, especially for precise low-activity measurements or multi-isotope discrimination.

\begin{figure}[t]
  \centering
  \begin{subfigure}[b]{0.48\linewidth}
    \includegraphics[width=\linewidth]{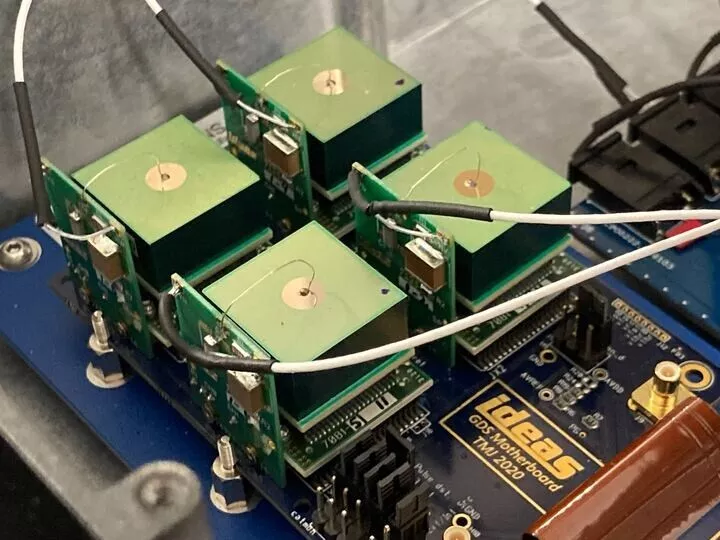}
  \end{subfigure}\hspace{0.5em}
  \begin{subfigure}[b]{0.48\linewidth}
    \includegraphics[width=\linewidth]{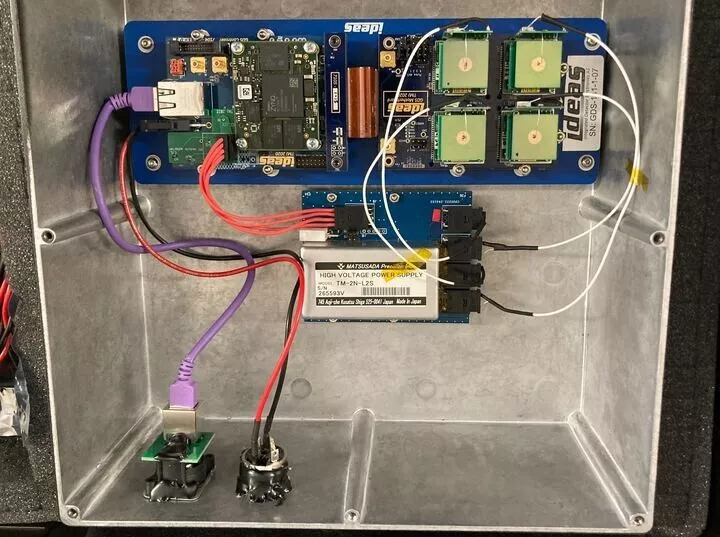}
  \end{subfigure}
  \caption{IDEAS GDS-100 hardware. Left: close-up of the $2\times2$ detector assembly comprising four CZT modules (each $11\times11$ channels), Right: full enclosure of the system.}
  \label{fig:gds100}
\end{figure}

Therefore, in this study, we present a CZT-based micro-activity dose calibrator using Monte Carlo simulation and compare its performance with a conventional NaI(Tl) well counter for low activity $^{225}$Ac measurements. 
The proposed system is benchmarked against a commercial  CZT detector (IDEAS GDS-100; Fig.~\ref{fig:gds100}) and is designed to improve activity quantification in the $\mu$Ci--nCi range relevant to TAT. 
Key innovations include configuration modification and the application of time-coincident pixel-level clustering method, aimed at increasing the net counts in full-energy-peak (FEP)–based activity estimation by recovering events otherwise lost to Compton scatter.

\section{Materials and Methods}
\subsection{Design of CZT-based micro-activity dose calibrator}
The proposed CZT-based micro-activity dose calibrator was benchmarked against a commercial CZT detector (IDEAS GDS-100) and was designed and evaluated using GATE (Geant4 Application for Tomographic Emission), a Monte Carlo simulation toolkit. Two techniques were implemented to improve activity quantification at very low activities: (i) configuration modification and (ii) time-coincident pixel-level clustering.

\begin{table}[ht]
\centering
\caption{Geometric properties of the commercial CZT detector (IDEAS GDS-100) used as a benchmark.}
\begin{tabular}{lll}
\hline
 & \textbf{CZT module} & \textbf{Detector assembly} \\
\hline
Array & $11\times 11$ (pixelated anode) & $2\times 2$ \\
Dimension & $1.22\times1.22~\mathrm{mm^2}$ (pixel) & $20 \times 20 \times 10~\mathrm{mm^3}$ (per module) \\
Sensitive volume & -- &  $\ge 16 ~\mathrm{cm^3}$ \\
Pitch & 1.72~mm (pixel pitch)& 32~mm \\
\hline
\end{tabular}
\label{tab:gds100_geometry}
\end{table}

\subsubsection{Configuration modification}
The detector configuration was changed from the original \emph{parallel} layout to a \emph{box-shaped well} configuration to increase the solid angle coverage and, consequently, the geometric efficiency.
The geometric efficiency is defined as
\begin{equation}
\varepsilon_{\mathrm{geom}} \;=\; \frac{N_{\mathrm{enter}}}{N_{\mathrm{emit}}},
\end{equation}
where $N_{\mathrm{emit}}$ is the number of emitted photons from the source volume and $N_{\mathrm{enter}}$ is the number of photons incident on the active CZT volume.
The geometric efficiency is well approximated by the solid angle fraction $f_{\Omega} \equiv \Omega/4\pi$ and the solid angle was evaluated via Monte-Carlo simulation. 

\paragraph{Solid-angle gain.}
For a point source positioned 0.5~cm off-center in the \emph{parallel} layout, the simulated solid angle was 
$\Omega_{\mathrm{parallel}} = 2.3988~\mathrm{sr}\ \pm 0.06\%$.
When the source was placed at the geometric center of the \emph{box-shaped well}, the solid angle increased to 
$\Omega_{\mathrm{well}} = 7.3328~\mathrm{sr}\ \pm 0.03\%$,
corresponding to a factor of $\Omega_{\mathrm{well}}/\Omega_{\mathrm{parallel}} \approx 3.06$ (i.e., $\sim$3.1$\times$).
This geometrical gain directly supports higher geometric efficiency and larger FEP net counts for activity estimation.

\subsubsection{Time-coincident pixel-level clustering}
\begin{figure}[t]
  \centering
  \begin{subfigure}[t]{0.52\linewidth}
    \centering
    \raisebox{1.8cm}{
      \includegraphics[width=\linewidth]{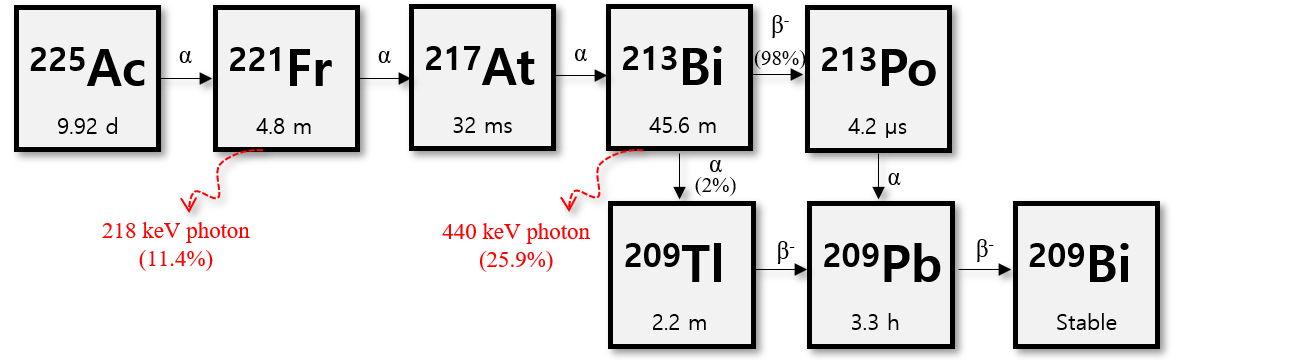}
    }
    \put(-240,140){\textbf{(a)}}
    \label{fig:ac225_decay_chain}
  \end{subfigure}\hfill
  \begin{subfigure}[t]{0.48\linewidth}
    \centering
    \includegraphics[width=\linewidth]{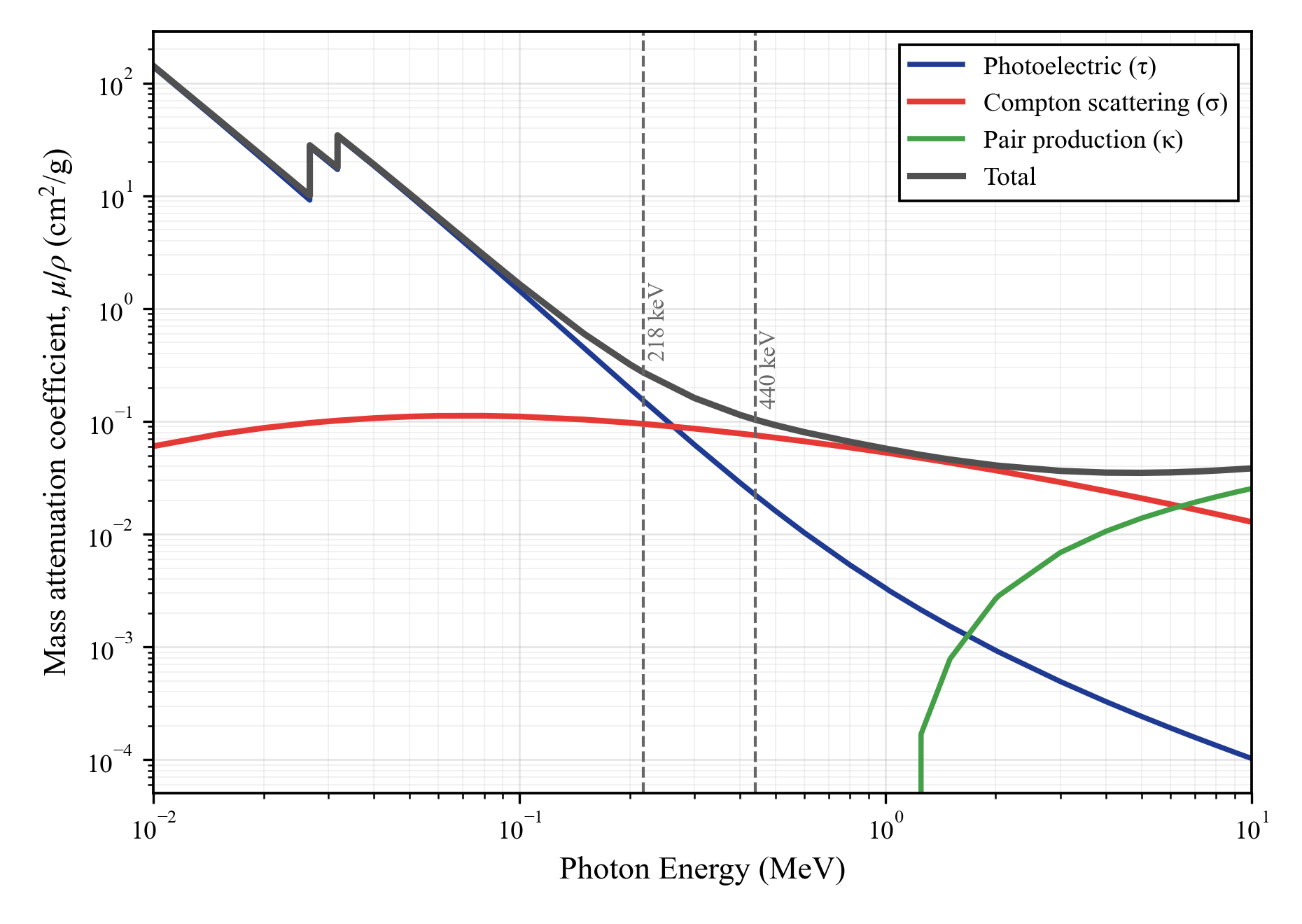}
    \put(-210,140){\textbf{(b)}}
    \label{fig:czt_interaction}
  \end{subfigure}
  \caption{(a) Decay chain of $^{225}$Ac with photon emissions at 218 and 440 keV, and (b) interaction processes in CZT with these emissions highlighted.}
  \label{fig:ac225_czt}
\end{figure}

With a monolithic CZT sensor employing a pixelated anode readout and compact geometry, a single photon often distributes its deposited energy across multiple anode pixels via charge sharing and multi-site Compton interactions, causing pixel-wise analyses to miss full-energy-peak (FEP) counts. This leads to a quantitative loss in full-energy-peak (FEP) counts when events are analyzed pixel-wise. 
Moreover, in CZT (with $Z_{\mathrm{eff}}\!\approx\!50$), Compton scattering becomes increasingly dominant above $\sim$300~keV. 
Notably, $^{225}$Ac—a promising $\alpha$-emitting radionuclide for targeted alpha therapy (TAT)—has prominent gamma emissions near 218 and 440~keV, placing the latter in a regime where Compton interactions are prevalent as shown in Fig.~\ref{fig:ac225_czt}.

To recover this loss, we applied a time-coincident pixel-level clustering that merges temporally and spatially proximate hits into a single event. Given a timing window $\Delta t$ and a spatial radius $R$, two hits $i$ and $j$ are considered adjacent if
\[
|t_i - t_j| \le \Delta t
\quad\text{and}\quad
d(i,j) \le R,
\]
where $d(i,j)=\lVert \mathbf{u}_i-\mathbf{u}_j\rVert_2$ with $\mathbf{u}_i=(x_i,y_i,z_i)$.

\paragraph{Inputs and parameters.}
For the time-ordered hit lists $H=\{(t_i,E_i,x_i,y_i,z_i)\}_{i=1}^N$,
two user parameters control the clustering:
(i) the coincidence window $\Delta t$ (ns) and
(ii) the spatial radius $R$ (mm). For a detector with pixel pitch $p$ (mm), we also defined the dimensionless radius $r \equiv R/p$ (“pixel-pitch multiples”; $r{=}1$ corresponds to one pixel pitch).

\paragraph{Clustering.}
Time-ordered hits are partitioned into non-overlapping, left-anchored timing windows of width $\Delta t$,
\[
W_i=\{\,k\ge i \mid t_k-t_i\le\Delta t\,\}, \qquad j=\max W_i.
\]

Within each window \(W_i\), we build an undirected \(\varepsilon\)-neighborhood graph \(G_{W_i}\) on the vertex set \(W_i\), connecting \((k,l)\) if
\[
\lVert \mathbf{u}_k - \mathbf{u}_l \rVert_2 \le R \qquad (k,l\in W_i),
\]
where \(\mathbf{u}_k=(x_k,y_k,z_k)\) denotes the hit position. 
The connected components \(\{\mathcal{C}_\ell\}_\ell\) of \(G_{W_i}\) define event-level clusters with energies
\[
E_{\mathrm{cluster}}(\mathcal{C}_\ell)=\sum_{k\in\mathcal{C}_\ell} E_k .
\]
A singleton window (\(|W_i|=1\)) yields one cluster with energy \(E_i\). We then advance to the first hit outside the current window \((i\leftarrow j+1)\) and repeat until \(i>N\). Merging the clusters from all windows yields the post-clustering spectrum \(S_{\mathrm{post}}\).

\paragraph{Random (delayed) surrogate for accidentals.}
We estimate accidentals at the same per-module singles rates—without true time correlation—using a per-face circular time roll. Let $f$ index detector faces (modules) and $\mathcal{I}_f$ be the hit indices on face $f$ with $n_f=|\mathcal{I}_f|$. For each $f$ with $n_f\!\ge\!10$, we roll the timestamp sequence by
\[
k_f=\bigl\lfloor \rho\,n_f \bigr\rfloor,\qquad \rho\simeq 0.41
\]
leaving $(x,y,z,E)$ unchanged and wrapping indices circularly. This constructs a time-decorrelated surrogate $\hat H$ that preserves marginal rates and dead-time structure within each face while destroying coincidences on $\mathcal{O}(\Delta t)$ scales. The same clustering applied to $\hat H$ yields a surrogate spectrum $S_{\text{rand}}$. We repeat the roll $N_{\text{roll}}$ times and average the resulting metrics to stabilize the accidental estimate.

\paragraph{Metrics.}
We derive and compare two metrics from the spectra—(i) \emph{true gain} and (ii) \emph{sideband continuum}—to quantify the effect of clustering.
Net counts around the photopeak energy $E_0$ are obtained within a narrow ROI by fitting a Gaussian plus linear background and integrating the Gaussian component.
Let $N_{\text{pre}}=\mathrm{net}(S_{\text{pre}})$ be the net counts from unclustered singles,
$N_{\text{post,raw}}=\mathrm{net}(S_{\text{post}})$ from clustered data, and
$N_{\text{rand}}=\mathrm{avg}\bigl[\mathrm{net}(S_{\text{rand}})\bigr]$ from the surrogate rolls.
We define the random-corrected post value as
\[
N_{\text{post,corr}} \;=\; N_{\text{post,raw}} - N_{\text{rand}} \, .
\]

\noindent\textbf{(i) True gain.}
To quantify FEP recovery, we report the percent increase relative to singles:
\[
\boxed{\text{true gain (\%)} \;=\; 100\,\frac{N_{\text{post,corr}}-N_{\text{pre}}}{N_{\text{pre}}}} .
\]

\noindent\textbf{(ii) Sideband continuum.}
Let $\sigma_{\text{pre}}$ be the Gaussian width from the \emph{pre}-spectrum fit.
We form symmetric sidebands referenced to $\sigma_{\text{pre}}$:
\[
\mathcal{D}_{\text{SB}}
= \frac{C\!\left([E_0-12\sigma_{\text{pre}},\,E_0-6\sigma_{\text{pre}}]\right)
      + C\!\left([E_0+6\sigma_{\text{pre}},\,E_0+12\sigma_{\text{pre}}]\right)}
     {12\sigma_{\text{pre}}} \quad \text{[counts/keV]},
\]
where $C([\cdot])$ counts events in the interval and the denominator yields a per-keV density.
We report the fractional change
\[
\boxed{\Delta \mathcal{D}_{\text{SB}}(\%) \;=\;
100\,\frac{\mathcal{D}_{\text{SB,post}}-\mathcal{D}_{\text{SB,pre}}}
               {\mathcal{D}_{\text{SB,pre}}}} ,
\]
with “post” computed on clustered energies and “pre” on unclustered singles.

\begin{figure}[t]
  \centering
  \begin{subfigure}[t]{0.8\linewidth}
    \centering
    \begin{tikzpicture}
      \node[inner sep=0] (imgA) {\includegraphics[width=\linewidth]{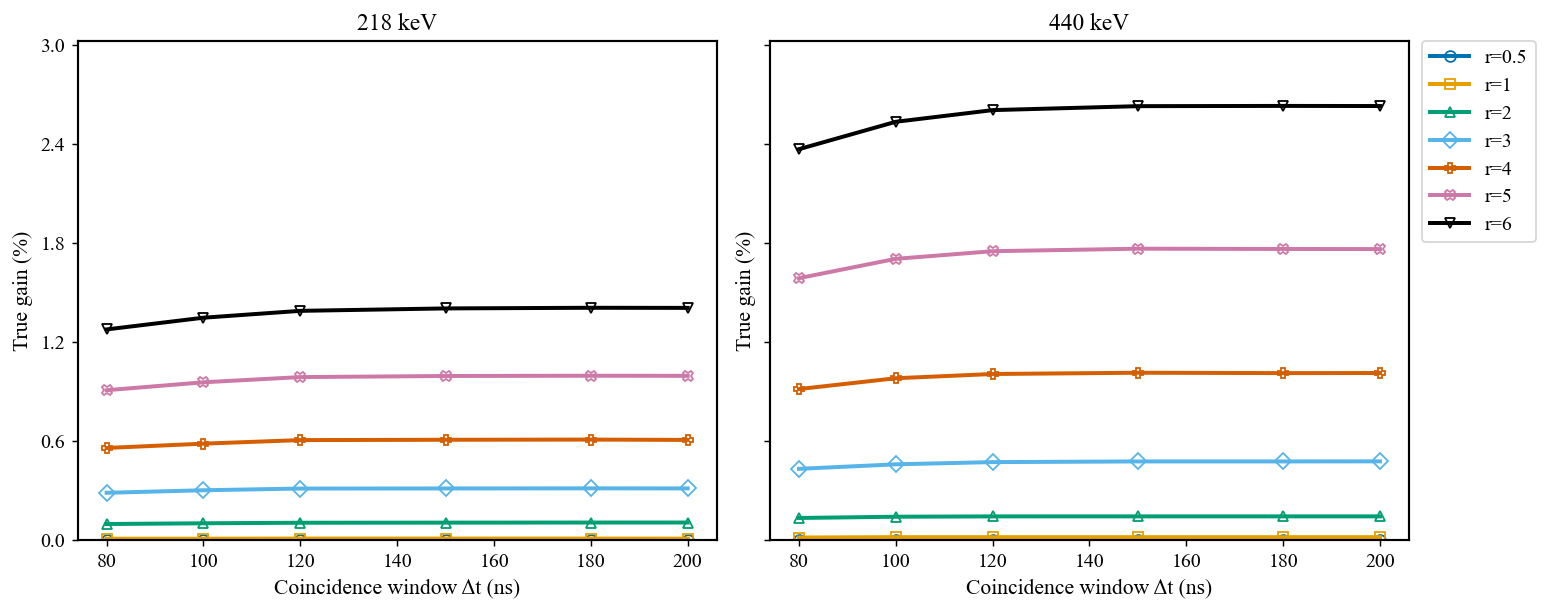}};
      \node[anchor=north west, xshift=-9pt, yshift=-4pt, font=\bfseries\normalsize] at (imgA.north west) {(a)};
    \end{tikzpicture}
    \phantomcaption\label{fig:true-gain}
  \end{subfigure}
  \begin{subfigure}[t]{0.8\linewidth}
    \centering
    \begin{tikzpicture}
      \node[inner sep=0] (imgB) {\includegraphics[width=\linewidth]{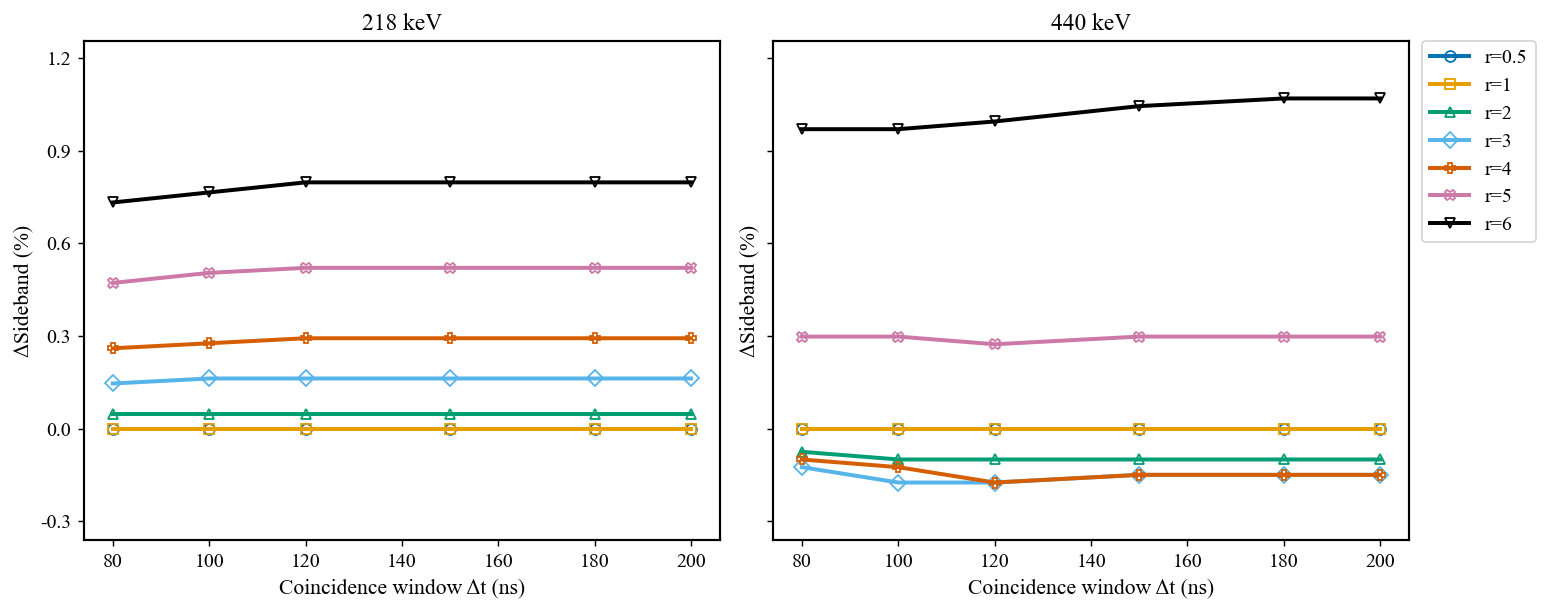}};
      \node[anchor=north west, xshift=-9pt, yshift=-4pt, font=\bfseries\normalsize] at (imgB.north west) {(b)};
    \end{tikzpicture}
    \phantomcaption\label{fig:sideband}
  \end{subfigure}
  \caption{Clustering performance at 218 and 440~keV: (a) true gain and (b) sideband continuum as functions of $\Delta t$ and $r=R/p$.}
  \label{fig:truegain-sideband-vertical}
\end{figure}

\paragraph{Parameter Optimization.}
The timing window $\Delta t$ and the dimensionless spatial radius $r=R/p$ were optimized on simulation data generated independently for the 218 and 440~keV photopeaks via a grid search
\[
\Delta t \in \{80,100,120,150,180,200\}\,\mathrm{ns},\qquad
r \in \{0.5,1,2,3,4,5,6\},
\]
maximizing \emph{true gain} subject to a constraint on the \emph{sideband continuum}
.
As shown in Fig.~\ref{fig:truegain-sideband-vertical}, true gain increases mainly with $r$ and shows diminishing returns with $\Delta t$ beyond $\sim\!120$\,ns, whereas the sideband penalty grows primarily with $r$ and only mildly with $\Delta t$.
We therefore select $\Delta t = 120\,\mathrm{ns}$ and $r=4$ ($R=4p$) as the knee of the gain--background trade-off: at this operating point, the true gain is $0.606\%$ at $218$\,keV and $1.007\%$ at $440$\,keV, while the sideband continuum changes by $+0.293\%$ and $-0.174\%$, respectively.

\subsubsection{Monte Carlo simulation}

We used the Geant4 Application for Tomographic Emission (\texttt{GATE}), a Monte Carlo simulation toolkit, to model photon transport.
Particle transport employed the \texttt{emlivermore} physics list, which provides high–accuracy modelling of low–energy electromagnetic interactions. The Geant4 \texttt{RadioactiveDecay} process was enabled to simulate nuclide–specific emissions (e.g., the $^{225}$Ac decay chain).  An energy resolution of $0.8\%$ at $662~\mathrm{keV}$ as reported in the IDEAS module datasheet was simulated as a Gaussian blurring of the deposited energy value provided by \texttt{GATE}. A per–event timing resolution of $80~\mathrm{ns}$ (FWHM) and an energy threshold of 20~keV was applied per anode pixel.  No explicit pile–up or dead–time was considered since for extremely low activity sources it is a negligible effect.

\subsection{Design of NaI(Tl) well counter}
We implemented a NaI(Tl) well counter in \texttt{GATE} to emulate the commercial Hidex AMG $\gamma$ counter, which is equipped with a single 3\,inch well–type thallium-activated sodium iodide crystal.
For experimental benchmarking, measured spectra were used to calibrate the channel–to–energy relation and fit an energy–resolution curve.
The absolute full–energy–peak efficiency was derived from the $^{225}$Ac measurement data using the 218 and 440~keV photopeaks.

\section{Results and discussion}
\begin{figure}[t]
  \centering
  \includegraphics[width=0.85\linewidth]{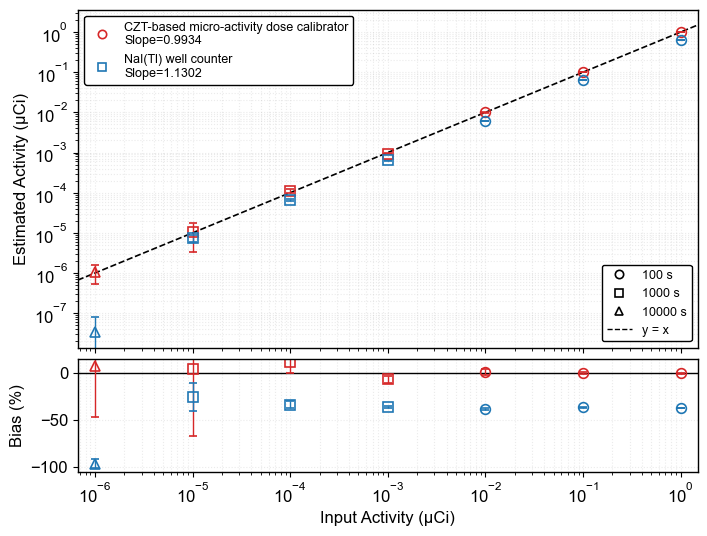}
  \caption{Accuracy and linearity for $^{225}$Ac using combined 218/440~keV FEPs. 
  CZT-based micro-activity dose calibrator (red) and NaI(Tl) well counter (blue); integration times: 100~s ($\circ$), 1000~s ($\square$), 10{,}000~s ($\triangle$). Fits shown are for \emph{Estimated} vs.\ \emph{Input}; dashed line denotes $y{=}x$.}
  \label{fig:accuracy}
\end{figure}

Performance evaluation followed the accuracy and linearity test described in NRC Information Notice 93-10. 
Serial dilutions of $^{225}$Ac from $1.0~\mu$Ci down to $10^{-6}~\mu$Ci (1~pCi) were measured with integration times of 100, 1000, and 10{,}000~s. 
Activities were reconstructed from the 218 and 440~keV full-energy peaks using Gaussian fits over a linear background with sideband subtraction, and the two peaks were combined by inverse-variance weighting (IVW).
Figure~\ref{fig:accuracy} shows estimated activity versus input activity (top) and the percent residuals (bottom).

The CZT-based micro-activity dose calibrator exhibited near-unity linearity across six orders of magnitude 
(slope $m=0.9934$, intercept $\approx 0$ for \emph{Estimated} vs.\ \emph{Input}), with percent-level bias over the entire range. 
Residuals remained within a few percent down to $10^{-6}~\mu$Ci (1~pCi), with a small positive bias at the lowest activity for the longest integration (10{,}000~s).

By contrast, the NaI(Tl) well counter remained linear but systematically under-responded relative to the identity line $y{=}x$ 
, consistent with well-geometry summing and resolution-driven ROI leakage into the Compton continuum. 
This under-response manifests as \(-20\%\) to \(-40\%\) bias over much of the range, and at $10^{-6}~\mu$Ci with 10{,}000~s acquisition the NaI(Tl) point approaches or falls below its quantitation capability, yielding a large negative residual.

\section{Conclusion}
We presented a CZT-based micro-activity dose calibrator tailored to the $\mu$Ci--nCi regime of $^{225}$Ac relevant to TAT.
Combining a box-shaped well geometry with time-coincident pixel-level clustering, the system achieved near-unity linearity with percent-level bias from $1.0$ to $10^{-6}\,\mu$Ci across the tested range.
A grid search identified $\Delta t=120$\,ns and $r{=}4$ ($R=4p$) as a robust operating point that improves FEP recovery while keeping sideband changes small.
Future work will include hardware validation across additional radionuclides and a comprehensive uncertainty budget.

\appendix

\acknowledgments
This work was supported by the Korea Institute of Energy Technology Evaluation and Planning (KETEP) and the Ministry of Trade, Industry \& Energy (MOTIE) of the Republic of Korea (RS-2023-00304393).


\bibliographystyle{JHEP}
\bibliography{biblio}

\providecommand{\href}[2]{#2}\begingroup\raggedright\begin{thebibliography}{10}

\bibitem{Intro1}
F.D.~Guerra~Liberal, J.M.~O'Sullivan, S.J.~McMahon and K.M.~Prise, \emph{Targeted alpha therapy: Current clinical applications}, \href{https://doi.org/10.1089/cbr.2020.3576}{\emph{Cancer Biotherapy and Radiopharmaceuticals} {\bfseries 35} (2020) 404}.

\bibitem{Intro2}
T.A.T.W.~Group, \emph{Targeted alpha therapy, an emerging class of cancer agents: A review}, \href{https://doi.org/10.1001/jamaoncol.2018.4044}{\emph{JAMA Oncology} {\bfseries 4} (2018) 1765}.

\bibitem{Intro3}
Y.S.~Kim and M.W.~Brechbiel, \emph{An overview of targeted alpha therapy}, \href{https://doi.org/10.1007/s13277-011-0286-y}{\emph{Tumor Biology} {\bfseries 33} (2012) 573}.

\bibitem{Intro4}
M.R.~Zalutsky and D.D.~Bigner, \emph{Radioimmunotherapy with $\alpha$-particle emitting radioimmunoconjugates}, {\emph{Acta Oncologica} {\bfseries 35} (1996) 373}.

\bibitem{Intro5}
M.~Zalutsky, J.~Schuster, P.~Garg, G.~Archer~Jr, M.~Dewhirst and D.~Bigner, \emph{Two approaches for enhancing radioimmunotherapy:$\alpha$ emitters and hyperthermia}, {\emph{Systemic Radiotherapy with Monoclonal Antibodies: Options and Problems} (1996) 101}.

\bibitem{Intro6}
T.K.~Nikula, M.R.~McDevitt, R.D.~Finn, C.~Wu, R.W.~Kozak, K.~Garmestani et~al., \emph{Alpha-emitting bismuth cyclohexylbenzyl dtpa constructs of recombinant humanized anti-cd33 antibodies: pharmacokinetics, bioactivity, toxicity and chemistry}, {\emph{Journal of Nuclear Medicine} {\bfseries 40} (1999) 166}.

\bibitem{Intro7}
Y.~Liu, T.~Watabe, K.~Kaneda-Nakashima, K.~Ooe, Y.~Shirakami, A.~Toyoshima et~al., \emph{Preclinical evaluation of radiation-induced toxicity in targeted alpha therapy using [211at] naat in mice: A revisit}, \href{https://doi.org/10.1016/j.tranon.2020.100757}{\emph{Translational Oncology} {\bfseries 13} (2020) }.

\bibitem{Intro8}
T.~Watabe, K.~Kaneda-Nakashima, K.~Ooe, Y.~Liu, K.~Kurimoto, T.~Murai et~al., \emph{Extended single-dose toxicity study of [211at]naat in mice for the first-in-human clinical trial of targeted alpha therapy for differentiated thyroid cancer}, \href{https://doi.org/10.1007/s12149-021-01612-9}{\emph{Annals of Nuclear Medicine} {\bfseries 35} (2021) 702 }.

\bibitem{Intro9}
S.~Farzipour, Z.~Shaghaghi, M.~Raeispour, M.~Alvandi, F.~Jalali and A.~Yazdi, \emph{Evaluation the effect of chelating arms and carrier agents in radiotoxicity of tat agents.}, \href{https://doi.org/10.2174/1874471015666220510161047}{\emph{Current radiopharmaceuticals} (2022) }.

\bibitem{Intro10}
A.~Bidkar, L.~Zerefa, S.~Yadav, H.~VanBrocklin and R.~Flavell, \emph{Actinium-225 targeted alpha particle therapy for prostate cancer}, \href{https://doi.org/10.7150/thno.96403}{\emph{Theranostics} {\bfseries 14} (2024) 2969 }.

\bibitem{Intro11}
A.~Morgenstern, C.~Apostolidis, C.~Kratochwil, M.~Sathekge, L.~Królicki and F.~Bruchertseifer, \emph{An overview of targeted alpha therapy with 225actinium and 213bismuth}, \href{https://doi.org/10.2174/1874471011666180502104524}{\emph{Current Radiopharmaceuticals} {\bfseries 11} (2018) 200 }.

\bibitem{Intro12}
H.~Song, M.~Xu, J.~Cai, J.~Chen, Y.~Liu, Q.~Su et~al., \emph{225ac-labeled antibody for fibroblast activation protein-targeted alpha therapy}, \href{https://doi.org/10.1021/cbmi.3c00067}{\emph{Chemical \& Biomedical Imaging} {\bfseries 1} (2023) 628 }.

\bibitem{Intro13}
A.~Juzeniene, V.~Stenberg, O.~Bruland and R.~Larsen, \emph{Preclinical and clinical status of psma-targeted alpha therapy for metastatic castration-resistant prostate cancer}, \href{https://doi.org/10.3390/cancers13040779}{\emph{Cancers} {\bfseries 13} (2021) }.

\bibitem{Intro21}
M.R.~McDevitt, D.~Ma, L.T.~Lai, J.~Simon, P.~Borchardt, R.K.~Frank et~al., \emph{Tumor therapy with targeted atomic nanogenerators}, \href{https://doi.org/10.1126/science.1064126}{\emph{Science} {\bfseries 294} (2001) 1537} [\href{https://arxiv.org/abs/https://www.science.org/doi/pdf/10.1126/science.1064126}{{\ttfamily https://www.science.org/doi/pdf/10.1126/science.1064126}}].

\bibitem{Intro19}
{\v{S}}vec and Schrader, \emph{An ionization chamber as a secondary standard for activity}, \href{https://doi.org/10.1016/S0969-8043(99)00222-5}{\emph{Applied radiation and isotopes: including data, instrumentation and methods for use in agriculture, industry and medicine} {\bfseries 52 3} (2000) 633}.

\bibitem{Intro20}
S.~Adler and P.~Choyke, \emph{Design and performance of the micro-dose calibrator}, \href{https://doi.org/10.1088/1361-6560/aada2f}{\emph{Phys. Med. Biol.} {\bfseries 63} (2018) 185004}.

\end{thebibliography}\endgroup

\end{document}